\documentclass[reprint,amsmath,amssymb]{revtex4-1}

\usepackage{graphicx}
\usepackage{amsmath}
\usepackage{dcolumn}
\usepackage{bm}
\usepackage{xcolor}

\begin{document}

\title{Laser-Ion Lens and Accelerator}

\author{Tianhong Wang$^1$, Vladimir  Khudik$^1$$^,$$^2$, and Gennady Shvets$^1$}
 \affiliation{$^1$School of Applied and Engineering Physics, Cornell University, Ithaca, New York 14850, USA.\\$^2$Department of Physics and Institute for Fusion Studies, The University of Texas at Austin, Austin, Texas 78712, USA.}

\date{\today}%

\begin{abstract}

Generation of highly collimated monoenergetic relativistic ion beams is one of the most challenging and promising areas in ultra-intense laser-matter interactions because of the numerous scientific and technological applications that require such beams. We address this challenge by introducing the concept of laser-ion lensing and acceleration (LILA). Using a simple analogy with a gradient-index lens, we demonstrate that simultaneous focusing and acceleration of ions are accomplished by illuminating a shaped solid-density target by an intense laser pulse at $\sim 10^{22}$W/cm$^2$ intensity, and using the radiation pressure of the laser to deform/focus the target into a cubic micron spot. We show that the LILA process can be approximated using a simple deformable mirror model, and then validate it using three-dimensional particle-in-cell simulations of a two-species plasma target comprised of electrons and ions. Extensive scans of the laser and target parameters identify the stable propagation regime where the Rayleigh-Taylor (RT)-like instability is suppressed. Stable focusing is found at different laser powers (from few- to multi-petawatt). Focused ion beams with the focused density of order $10^{23}$cm$^{-3}$, energies in access of $750$MeV, and energy density up to $2\times10^{13}$J/cm$^3$ at the focal point are predicted for future multi-petawatt laser systems.

\end{abstract}

\maketitle


{\it Introduction and Motivation.} A focusing optical lens is one of the oldest and best-known scientific instruments. The operating principle of a lens can be easily understood in either wave or corpuscular description of light: by causing a photon impinging on its central portion travel longer distance than a photon impinging on its periphery, we can ensure that both photons reach the focal point at the same time. Thus, focusing is ensured by the judicious variation of the lens thickness: thicker at the center, thinner at the edge. While the speed of the photons is piecewise constant inside and outside the lens, this is not a necessary condition for light focusing. For example, in a gradient index (GRIN) lens~\cite{GRIN_Book} the light speed continuously varies across the lens, thus ensuring that all photons arrive at the focal point at the same time, regardless of their entry point. Motivated by the concept of a GRIN lens focusing light using non-uniform matter, we pose the following question: is it possible to focus matter using light?

\begin{figure}[t!]
\centering
  \includegraphics[width=0.90\columnwidth]{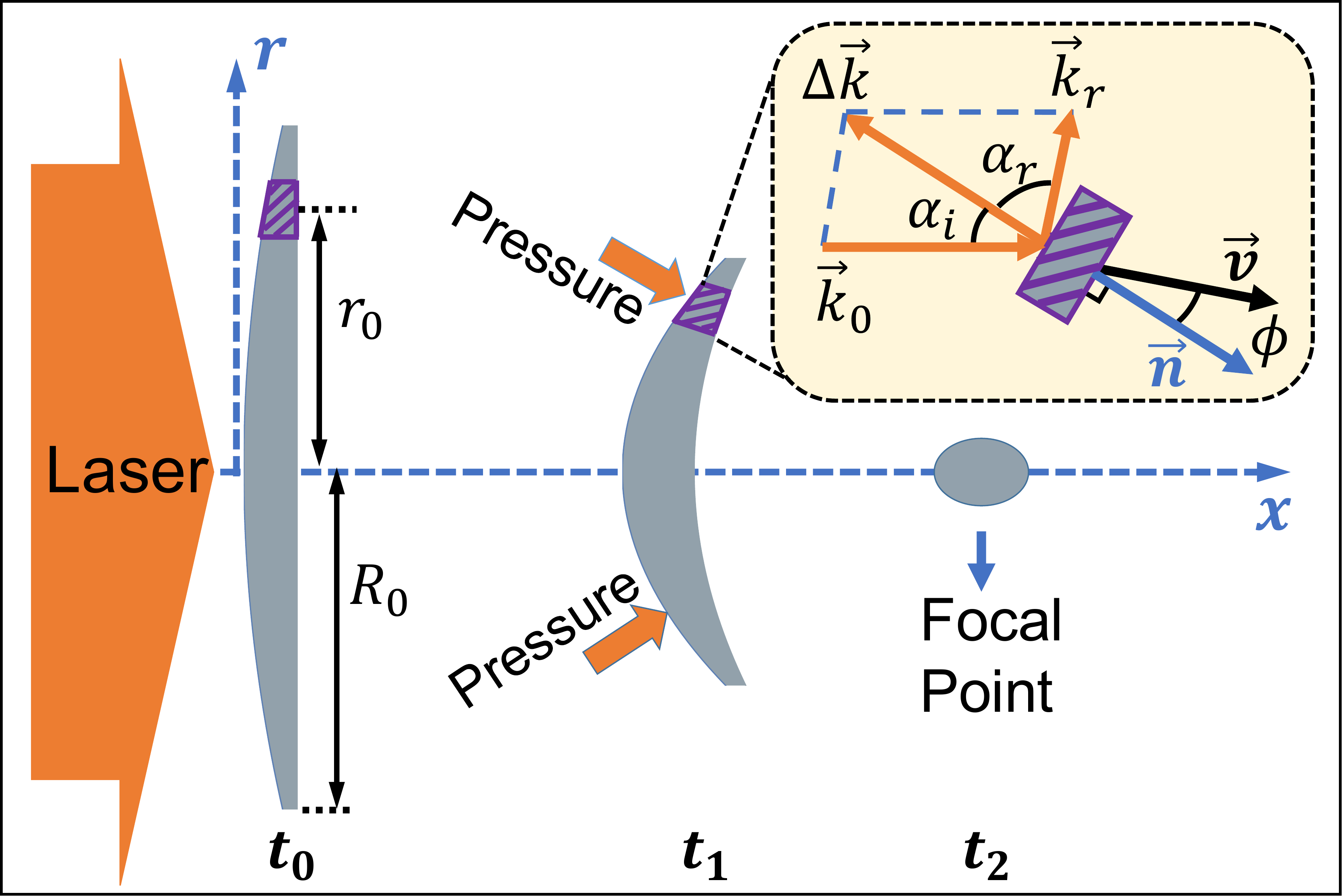} \\
  \caption{Schematic of the LILA concept: a laser beam propels a thin dense target with nonuniform thickness. Inset: the geometry of the laser reflection from a small target element moving with velocity $\vec{v}$.}\label{Fig1}
\end{figure}

The key to developing such a "matter lens" is the realization that, just as matter can change the velocity/direction of a photon, an intense flux of photons can do the same for the matter. This can be accomplished using the concept of radiation pressure acceleration (RPA) \cite{macchi2005RPA_1,pegoraro2007RPA_2,robinson2008RPA_3,qiao2009RPA_4} developed in the context of laser-ion acceleration of thin targets. The idea is schematically illustrated in Fig.~\ref{Fig1}, where the target is shaped in such a way that its outer (thinner) regions are accelerated to higher velocities than its central (thicker) region. We analytically demonstrate that, for a judicious choice of target areal density distribution, the resulting continuous velocity variation across the target enables its focusing into an infinitesimal spot. The important feature of RPA-based focusing of the matter is that the target is not only focused, but also accelerated. Hence, we refer to this scheme as a Laser-Ion Lens and Accelerator (LILA).

Just as the wave nature of light prevents its focusing to a geometric point by the GRIN lens, several fundamental plasma effects impose limits on the minimal focal spot of a realistic laser-propelled target. Those effects include Coulomb explosion \cite{fourkal2005coulomb} and Rayleigh-Taylor (RT)-like instability \cite{pegoraro2007RPA_2,robinson2008RPA_3,pukhov_prl10,khudik2014analytic_1D} that are known to break up constant-thickness targets, as well as plasma heating by the laser pulse. Under a simplifying assumption about the target as an initially cold two-species (electrons and single-charge ions) plasma, we describe the results of our fully-kinetic particle-in-cell (PIC) simulations and demonstrate that the RT-like instability and Coulomb explosion are effectively suppressed in a converging flow of the plasma. The result of the LILA mechanism is a quasi-monoenergetic and nearly-neutral relativistic beam which is both tightly-focused and (due to its ultra-low emittance) weakly-divergent. Scientific and industrial applications of such beams are wide-ranging: fast ignition of fusion targets \cite{tabak1994ignition,atzeni2002ignition_2}, production of warm dense matter \cite{patel2003WDM_1,dyer2008WDM_2}, hadron cancer therapy \cite{bulanov2002therapy_1,yogo2009therapy_2,bolton2010therapy_3}, and particle nuclear physics \cite{mckenna2005spallation_1,hannachi2007prospects_nuclear}.

The LILA concept owes its feasibility to recent advances in solid-state laser technology that have enabled the generation of ultra-short laser pulses with intensities well above $I_{\rm rel} = 1.37\times10^{18}$W/cm$^{-2}$~\cite{perry1999petawatt} corresponding to the normalized vector potential $a_0 \equiv eA/m_ec \sim 1$ for the laser wavelength $\lambda_0 \equiv 2\pi c/\omega_0 = 1 \mu$m, where $A$ is the laser vector potential, $c$ is the speed of light, $-e$ and $m_e$ are the electron charge and mass, and $\omega_0$ is the laser frequency. Because of the wide range of their applications, laser-driven ion accelerators represent one of the most exciting areas of plasma physics at high energy density. In addition to the RPA regime, where an over-dense thin target is propelled by the radiation pressure $P=2I/c$ of a circular polarized laser with ultra-high intensity $I > 10^{21}W/cm^{2}$ \cite{esirkepov2004Model_1D_efficiency,bulanov2010unlimitedModel_1D,yan2008spectrum_1,khudik2014analytic_1D}, several compelling ion accelerating scenarios are currently under investigation. Those include target normal sheath acceleration (TNSA) \cite{snavely2000TNSA_1,hatchett2000TNSA_2}, shock wave \cite{silva2004shocks_1,PhysRevLett.101.164802shocks_2}, and laser break-out afterburner (BOA) \cite{yin2006BOA_1,yin2006BOA_1} acceleration.

Because the emphasis of this work is on simultaneous acceleration and focusing of the target to a wavelength-scale focal spot, in the rest of the manuscript we concentrate on the RPA approach. While it is possible that under some conditions other ion acceleration technique could also provide focusing, their investigation is beyond the scope of this manuscript. As the starting point, we develop a model describing the dynamics of a laser-propelled deformed thin target under a simplifying assumption that the target acts on the incident laser light as a perfectly-reflecting mirror. This model is used to derive the optimal target shape enabling the ideal focusing of the target into a focal point.

{\it Deformable mirror model of LILA.} Interaction of a circularly polarized planar laser wave with a thin dense target, whose thickness $d(r_0)$ decreases from the target center ($r_0=0$) toward its edge ($r_0 = R_0$), can be simplified by modeling the target as an ideal mirror deformed during its motion by slowly-changing radiation pressure $P$ applied normally to the target surface. Because of the variation of the areal mass $dm/dS = m_i n_0 d$ (where $m_i$ is the ion mass and $n_0$ is the target density), different parts of the target experience different accelerations. The initially flat target bends forward because of the higher velocity of its periphery and is eventually focused by the applied radiation pressure to a small area. The evolving shapes of the target at different moments in time are schematically shown in Fig.~\ref{Fig1}. Despite the simplicity of the deformable mirror (DM) model, which neglects many plasma phenomena such as laser heating of the target~\cite{BrunleHeating,BrunleHeating2} and spatial separation between light electrons and heavy ions~\cite{khudik2014analytic_1D}, it is found useful for predicting the optimal thickness profile $d(r_0)$ and for deriving scaling laws of target's focusing and acceleration.

Assuming that an initially planar target starts out, and remains, axially-symmetric, we use the Lagrangian coordinates~\cite{munson2014fundamentals} to describe the motion of ring-shaped elements of the target. The two coordinates, $x(r_0,t)$ and $r(r_0,t)$, are functions of the time $t$ and the initial radial position $r_0$: $x(r_0,t=0)=0$ and $r(r_0,t=0)=r_0$.  The number of the ions $\delta N$ contained in a ring element of the width $\delta r_0$ and radius $r_0$ is conserved during its motion: $\delta N = 2\pi n_0 d(r_0) r_0\delta r_0$. The element's area $\delta S(r_0,t)$ and the unit vector $\vec{n}(r_0,t)$ normal to the element's surface are changing with time according to
\begin{eqnarray}
    \delta S &=&2\pi r(r_0,t)[r'(r_0,t)^2+x'(r_0,t)^2]^{1/2}\delta r_0,\label{eq_N_S}\\
    \vec{n}&=&\frac{r'(r_0,t)\vec{e}_x-x'(r_0,t)\vec{e}_r}{[r'(r_0,t)^2+x'(r_0,t)^2]^{1/2}},\label{eq_n}
\end{eqnarray}
where $^{\prime}$ stands for a derivative with respect to $r_0$, and $(\vec{e}_x,\vec{e}_r)$ are the  unit vectors in the propagation and radial directions, respectively.

When photons are bounced from a perfectly-reflecting target moving with velocity $\vec{v}$, the reflection angle $\alpha_r$ differs from the incidence angle $\alpha_i$. However, the change of the photon momentum $\hbar\Delta \vec{k}$ is always directed along the surface normal because of the accompanying redshift of the photon frequency~\cite{gjurchinovski2004reflection,galli2012reflection}. After cumbersome but straightforward application of the momentum conservation, an equation of motion for the target element is obtained:
\begin{eqnarray}\label{eq_EOM}
    \frac{\delta N}{\delta S} \frac{\partial\vec{p}}{\partial t} &=& -\kappa \Big( \frac{E^2}{2\pi m_ic} \cos\alpha_i\Big) \frac{(\beta\cos\phi-\cos\alpha_i)}{(1-\beta^2\cos^2{\phi})} \vec{n},
\end{eqnarray}
where $\beta=v/c$ is the dimensionless velocity, $E$ is the laser electric field,  $\vec{p} = \vec{\beta}/\sqrt{1-\beta^2}$ is the dimensionless relativistic ion momentum, and $\kappa = (\cos\alpha_i-\beta\cos\phi)/\cos\alpha_i$.

\begin{figure}[h!]
\centering
  \includegraphics[width=0.98\columnwidth]{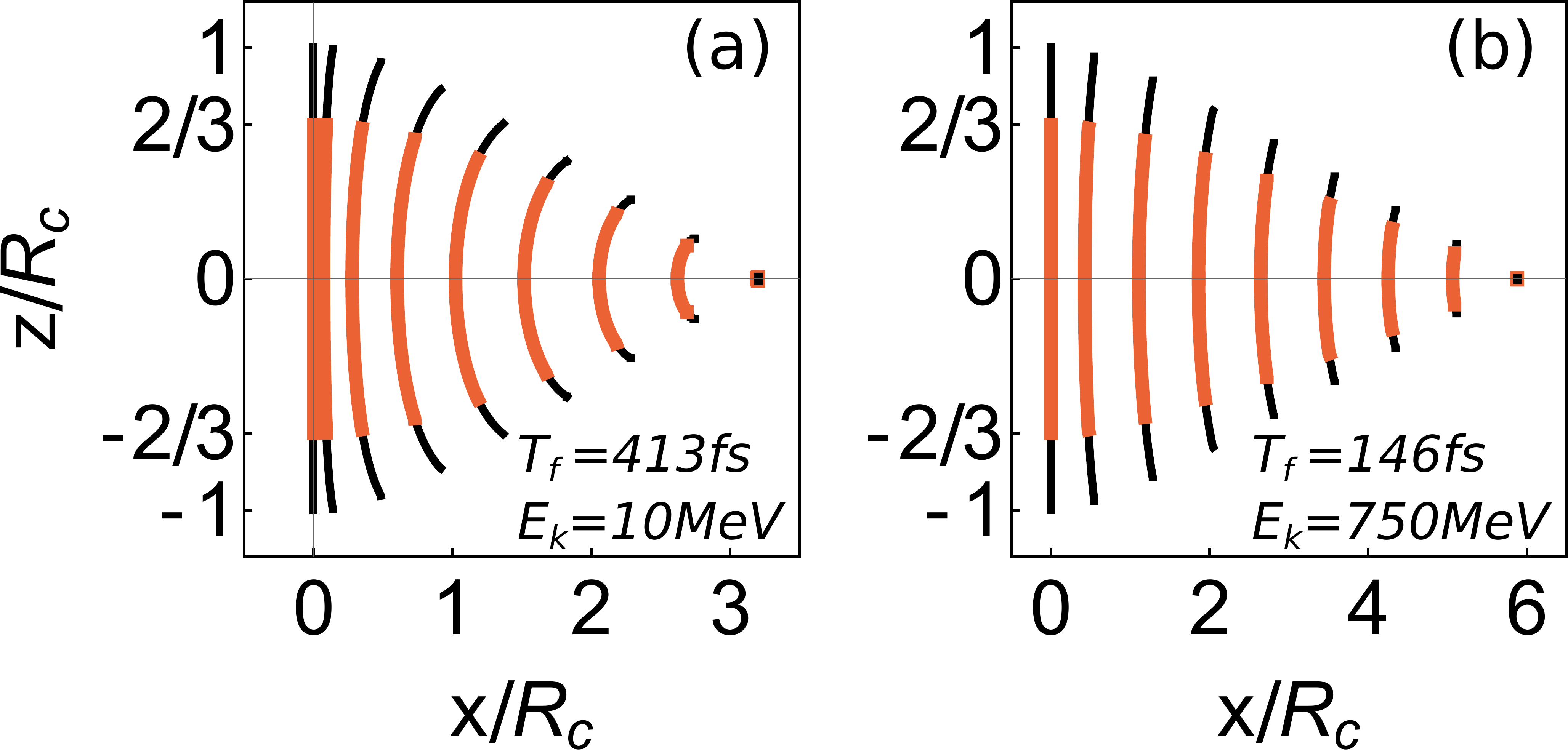}\\
\caption{Deformable Mirror (DM) model of the acceleration/focusing of a thin target propelled by laser pulses of different normalized amplitudes $a_0=10$ ($\Gamma=0.021$) (a) and $a_0=100$ ($\Gamma=2.1$) (b), and different target radius: $R_0=R_c$ (black curves) and $R_0=\frac{2}{3}R_c$ (red curves). Target parameters: $R_c=6\mu$m, $d_0=300 {\rm nm}$, and $n_0=100n_c$. }\label{Fig2}
\end{figure}

Using a geometric optics analogy with an aberations-free parabolic lens~\cite{hecht2016optics}, we consider a parabolically shaped target with radius $R_0$ and variable thickness given by $d(r_0) = d_0 (1 - r_0^2/2R_c^2)$, where  $d_0$ is the target thickness at the center and $R_c$ is the radius of curvature.  After normalizing the spatio-temporal coordinates according to $x \rightarrow x/R_c$, $r \rightarrow r/R_c$, $r_0 \rightarrow r_0/R_c$, $t \rightarrow ct/R_c$ by $R_c$, and $d \rightarrow d/d_0$, the target's equations of motion are expressed as
\begin{eqnarray}
\frac{\partial \vec{p}}{\partial t} &=& \frac{gR_c}{c^2} \frac{(\cos\alpha_i-\beta\cos\phi)^2}{(1-\beta^2\cos^2{\phi})} \frac{r/r_0}{d(r_0)} (r'\vec{e}_x-x'\vec{e}_r), \label{eq_EOM_nom} \\
\frac{\partial\vec{r}}{\partial t} &=& \frac{\vec{v}}{c} = \frac{\vec{p}}{(1+{|\vec{p}|}^2)^{1/2}},\label{eq_r}
\end{eqnarray}
where $g = E^2/2\pi d_0m_in_0$ is the initial acceleration of the central point of the target, and $d = (1 - r_0^2/2)$ is the normalized target thickness.
The trigonometric functions in Eq.~(\ref{eq_EOM_nom}) can be expressed as $\cos\phi = \vec{n}\cdot \vec{v}/v$ and $\cos\alpha_i = \vec{n} \cdot \vec{e}_x$, where $\vec{n}$ is given by Eq.~(\ref{eq_n}). Assuming an initially stationary target ($\vec{p}(r_0,t=0) = 0$ for all values of $r_0<R_0/R_c$, we observe that the target dynamics is determined by just two dimensionless parameters: the target radius $R_0/R_c$ and its peak energy $\Gamma \equiv gR_c/c^2$. The final target energy becomes relativistic for $\Gamma \sim 1$.

The results of the numerical solutions of the Eqs.(\ref{eq_EOM_nom},\ref{eq_r}) are presented in Figure~\ref{Fig2}, where several time snapshots of the target shape are shown. For each normalized laser amplitude ($a_0=10$ in (a) and $a_0=100$ in (b)) we simulated two initial target radius: $R_0=R_c$ (black lines) and $R_0 = 2R_c/3$ (red line). Highly over-dense $H^{+}-e^{-}$ plasma  with $n=100n_c$ was assumed inside the target, where $n_c \equiv m_e \omega_0^2 /4\pi e^2 = 1.12\times 10^{21}$cm$^{-3}$ is the critical density for the laser wavelength $\lambda_0 = 1 \mu$m. In all four cases, the parabolically shaped target is focused aberration-free to a very small spot at the focusing distance $x = L_{\rm f}$.

In the sub-relativistic case ($\Gamma=0.021$), the target undergoes significant bending, and its final (per proton) kinetic energy reaches $E_k \approx 10$MeV at the focal point $L_{\rm f} \approx 3 R_c$. In the relativistic case ($\Gamma=2.1$), the target bending is smaller, and  the final proton kinetic energy is $E_k \approx 750$MeV at $L_{\rm f} \approx 6R_c$. In fact, it can be analytically demonstrated (see supplemental material) that $L_f \approx 2.95 R_c$ in the sub-relativistic limit of $\Gamma \rightarrow 0$. In the relativistic  ($\Gamma \gg 1$) case, the focusing length $L_{\rm f}$ monotonically increases with $\Gamma$.  Another important observation from Fig.~\ref{Fig2} is that the focusing length is essentially independent of the initial target radius: both targets with different radii focus at the same point.  Therefore, within the limits of the DM model, the target dynamics is parametrized by $\Gamma$ alone.

In reality, the applicability of the DM model is limited by the complex dynamics of multi-species plasmas that includes plasma heating (which cannot be completely eliminated even for circular polarization because of the non-planar nature of the bending target), charge separation between electrons and ions, the Coulomb explosion that follows from such separation, and the RT-like instability. Below we demonstrate that, despite the complexity of relativistic laser-plasma interactions, the conclusions of the DM model largely hold, and that simultaneous focusing/acceleration by the LILA mechanism is indeed feasible under a wide range of laser powers.

{\it Particle-in-Cell Simulations of LILA.} We validate the LILA concept by performing three-dimensional (3D) simulations using a first-principles PIC code VLPL~\cite{pukhov1999VLPL3D}. In the first example, we assume a fully-ionized two-species (electrons and protons) thin parabolically shaped plasma target with $R_0=8\mu$m and $R_c = 7\mu$m, the initial electron/proton density $n=100n_c$, and a circular-polarized planar wave with wavelength $\lambda_0 = 1\mu m$ and intensity  $I = 1.75\times10^{22}W/cm^{2}$ ($a_0=80$). These parameters correspond to the estimated laser power $P = 35$PW over the target area. The target thickness $d_0=300 {\rm nm}$ at its center is chosen to be slightly larger than the optimal RPA thickness~\cite{tripathi2009_optimal_thick} $d_{\rm opt} = (\lambda/\pi)(n_c/n)a_0 \approx 250$nm.
\begin{figure}[htp!]
\centering
 \includegraphics[width=0.98\columnwidth]{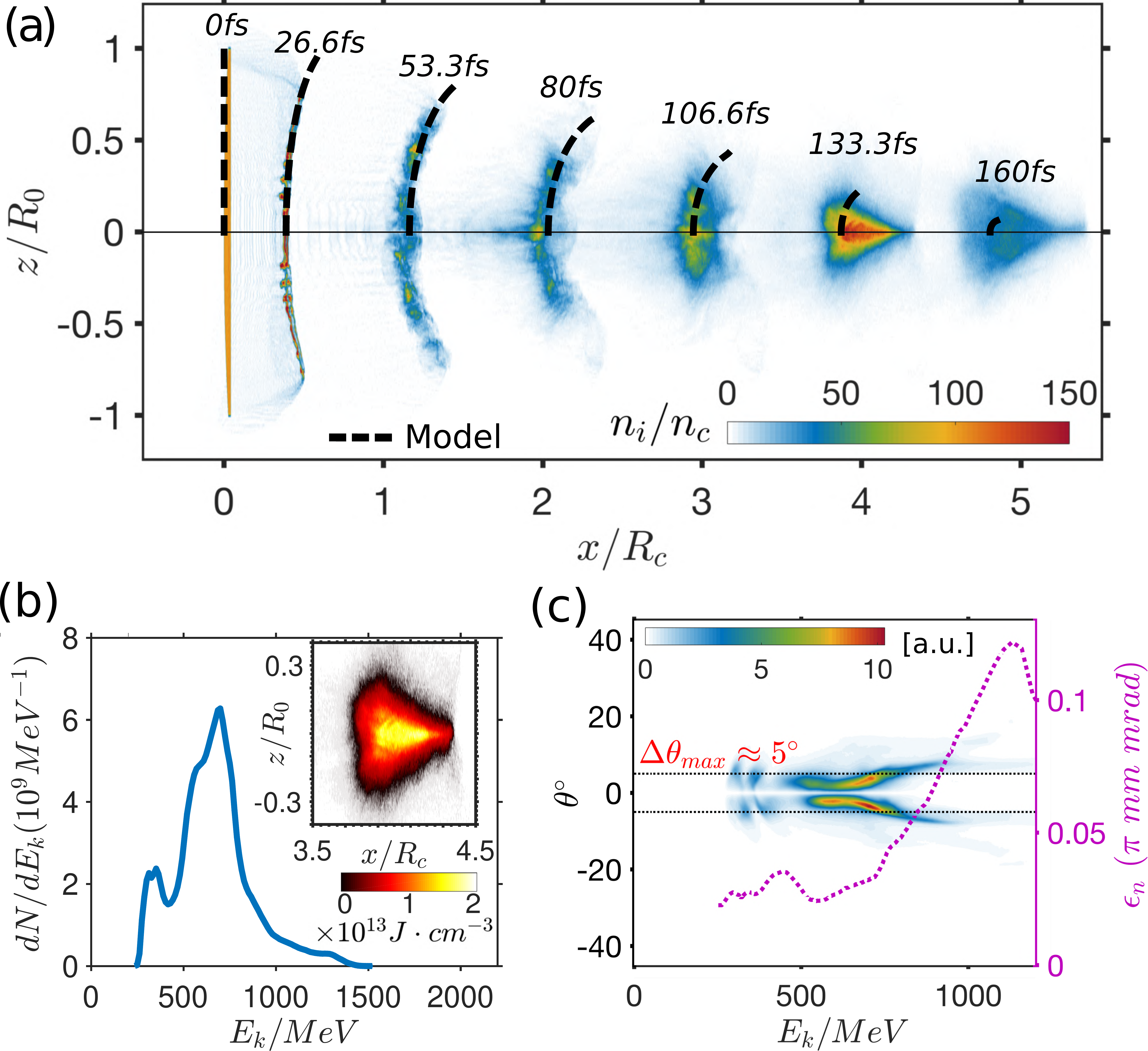}\\
\caption{A 3D PIC simulation of LILA. (a) Snapshots of ion densities. Black-dashed lines: target position from the DM model.  The focal spot (peak plasma density) is achieved at $t_f = 133.3$fs (b) Proton energy spectrum and energy density distribution (in the inset) at $t=t_f$. (c) Proton phase space ($E_k,\  \theta$) distribution and normalized emittance $\varepsilon_n$ (dotted line) vs energy $E_k$ at $t=t_f$. The simulation box size is $X\times Y\times Z=50\lambda\times20\lambda\times20\lambda$, consists with $5000\times250\times250$ cells. At the plasma region, each cell contains 160 macro-particles.}\label{Fig3}
\end{figure}

As shown in Fig.~\ref{Fig3}~(a), the time-dependent positions of the target (its bending and focusing) calculated using the VLPL code are very close to those obtained from the DM model. But in contrast to the DM model, the simulations predict target deterioration at the edges, where its thickness is smaller than $d_{\rm opt}$. Besides, we observe that a realistic plasma target cannot be focused into a point due to its stretching in the longitudinal ($x$-) dimension. Moreover, only a fraction of the ions ($\approx 50\%$) is focused into a focal spot measuring less than $4\mu$m in every dimension. It is well-known from flat target simulation that a considerable fraction of the ions is left in the tail of the target~\cite{khudik2014analytic_1D,tripathi2009_optimal_thick}, and that only some of the ions gain large energy through the RPA mechanism. Nevertheless, despite the target elongation and partial loss of ions, the peak density of the focused ions is $\approx 1.5$ times {\it larger} than their initial density due to the convergent plasma flow. Another deviation from the DM model is that the focal length $L_f^{({\rm PIC})} \sim 4 R_c \approx 28\mu$m found from the PIC simulations is slightly shorter than $L_f^{({\rm DM})} \sim 5 R_c=35\mu m$ predicted by the DM model.

Based on the value of $\Gamma\approx 1.6$ for the simulation parameters of Fig.~\ref{Fig3}, the target ions at the focal spot are expected to acquire relativistic energies. This is confirmed by the ion energy spectrum peaked at $E_k^{({\rm peak})} \approx 750$MeV plotted as a solid line in Fig.~\ref{Fig3}~(b). To quantify the degree of directionality of the LILA ions, we have plotted in Fig.~\ref{Fig3}~(c) the normalized  emittance $\epsilon_{n}(E_k) \equiv ({|\bf{p}}|/mc) \sqrt{\big< z^2 \big> \big< z'^2 \big> - \big< zz' \big>^2}$ as a function of ion energy $E_k$. Here $z' =p_z/p_x$, and the brackets $<\cdot\cdot\cdot>$ denote averaging over all particles with energies close to $E_k$. Remarkably, $\epsilon_{n}(E_k)$ has a minimum around $E_k \approx E_{\rm ion}^{({\rm peak})}$, indicating that the accelerated beam is not only focused and quasi-monoenergetic, but also highly directional.

Indeed, the proton beam distribution plotted in Fig.~\ref{Fig3}~(c) in the $(E_k,\ \theta)$ phase space (where $\theta$ is the angle between ion velocity and the $x$-axis) confirms that the angular spread of the peak energy ions at the focal spot is very small: $\Delta\theta_{max}\approx5^{\circ}$. This corresponds to the remarkably low emittance of quasi-monoenergetic ions at the focal spot: $\epsilon_{\rm min} \approx 0.035(\pi\cdot {\rm mm} \cdot {\rm mrads})$. This low emittance is preserved after the focal point, making the resulting ion beam interesting for a variety of applications that require collimated beams. The high concentration of relativistic ions in such a small focal volume results in an extremely high energy density $u_k$ plotted in the inset of Fig.~\ref{Fig3}~(b), with its peak reaching $u_k^{\rm max} \approx 2\times10^{13} {\rm J} \cdot$cm$^{-3}$.

\begin{figure}[htp!]
\centering
 \includegraphics[width=0.98\columnwidth]{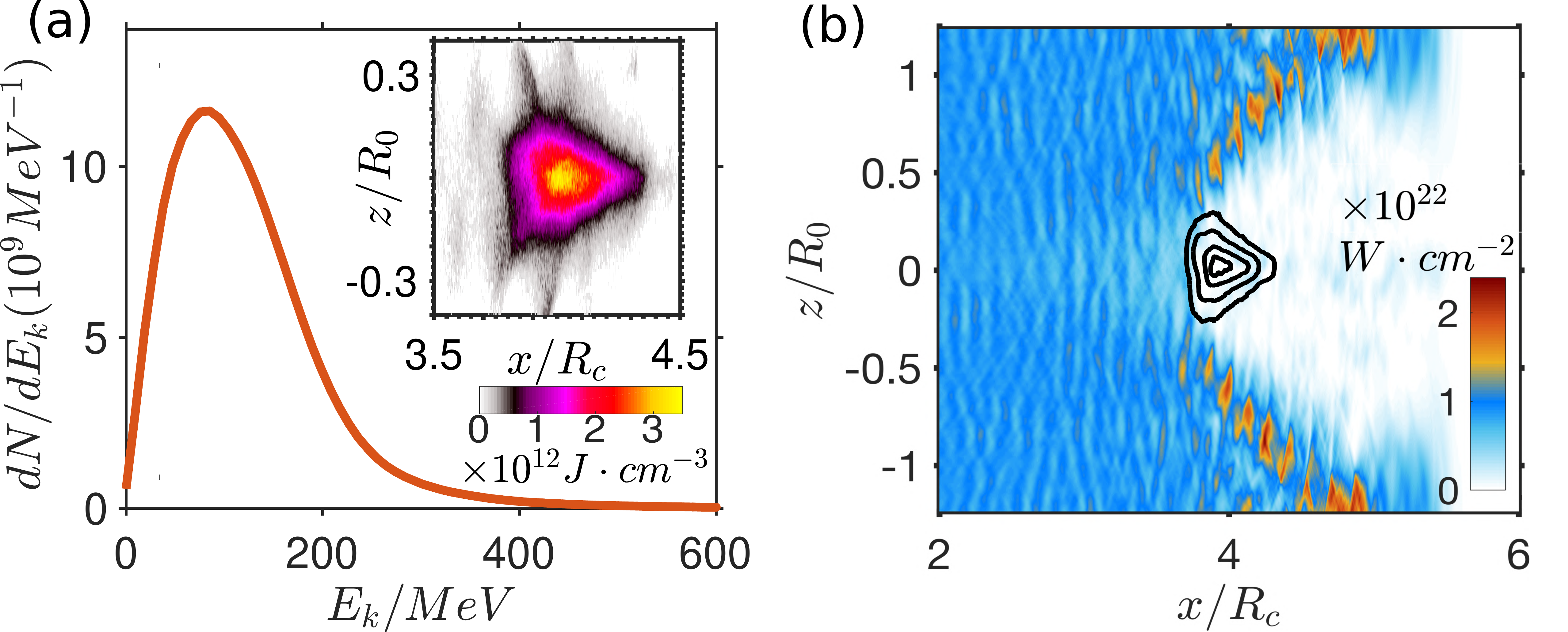}\\

\caption{(a) The electron energy spectrum (solid line) and energy density distribution (inset) at the focal spot. (b) Laser intensity distribution at $t = t_f$. Black electron density contours indicate the electron beam's location at $t = t_f$.}\label{Fig4}
\end{figure}

To understand why the DM model remains quite accurate in this regime of ultra-intense laser-matter interaction, we calculated the electron spectrum at $t=t_f$, as well as the spatial electron energy density distribution around the focal spot. We observe from  Fig.~\ref{Fig4}~(a) that the electrons at the focal spot remain significantly colder than the ions: their energy spectrum peaks at $E_{\rm el}^{\rm max} \sim 100$ MeV, and their peak energy density reaches only $u_{\rm el}^{\rm max} \approx 3.3\times 10^{12} {\rm J} \cdot$ cm$^{-3}$. Therefore, the two-species plasma target indeed behaves as quasi-neutral, with electrons moving slightly ahead of the ions to provide the charge separation needed to generate the requisite accelerating electric field.

We note that not only the laser pulse deforms and focuses the target, but the target itself strongly modifies the initially planar laser wavefront. The wavefront co-evolves with the target shape during its acceleration/focusing, and by the time $t = t_f$ it acquires a bowl-like shape shown in Fig.~\ref{Fig3}~(b). Note that only $\tau_L \approx 46.6$fs pulse length is needed until the focal point is reached. The laser energy $U_L$ contained within the corresponding volume $V_L = c\tau_L \times \pi R_0^2$ is $U_L \approx 1.6$kJ; a considerable fraction $\eta \approx 16\%$ of $U_L$ is transferred to the ions at the hot focal spot.

\begin{figure}[b!]
\centering
\includegraphics[width=0.98\columnwidth]{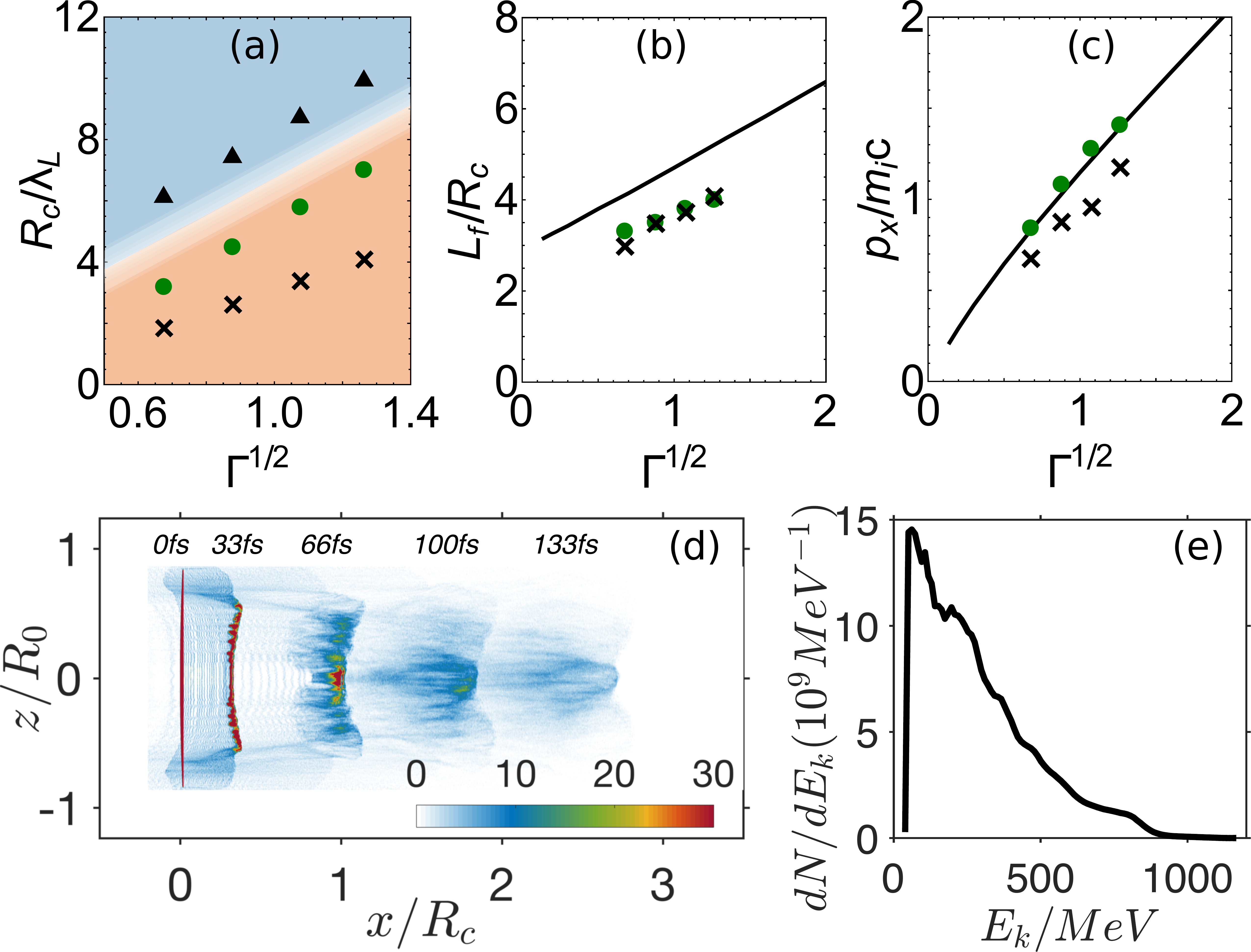}\\

\caption{ (a) Stable (crosses, bullets) and unstable (triangles) target focusing in the ($\Gamma,\, R_c$) parameter space. (b) Focal length $L_f$ and (c) momentum $p_x$ as functions of  $\sqrt{\Gamma}$  from the DM model (solid line) and 3D PIC simulations (crosses, bullets). (d) Example of the unstable acceleration/focusing of the target, corresponds to $\Gamma=1.6$ and $R_c/\lambda_L = 10$. (e) Ions energy spectrum of the unstable target in (d) at moment $t=133fs$.}\label{Fig5}
\end{figure}

{\it LILA scaling and stability.} With the DM model validated by 3D PIC simulation for at least some laser/target parameters, we next obtain simple scalings of the target's energy gain and focal distance that applies for a wide parameter range. As demonstrated earlier, the dynamics of the target focusing and acceleration within the DM model is determined by a single dimensionless parameter $\Gamma$. In particular (see supplemental material), the ion momentum $p_x$ at the focal point and the focusing length $L_f$ in this model can be approximated by the following expressions:
\begin{eqnarray}\label{eq_emp}
p_x/m_ic\approx\Gamma^{1/2},\quad
L_f/R_c\approx2\Gamma^{1/2}+2.95. \label{eq_f}
\end{eqnarray}
Approaching these scalings requires that the target does not succumb to RT-like instability. Therefore, we have carried out a series of VLPL simulations to examine the influence of the RT instability on the target focusing, and to verify the scalings given by Eq.~(\ref{eq_f}). The results of these simulations corresponding to the normalized quantities $\Gamma$ and $R_c/\lambda_L$ listed in Table I are shown in Fig.~\ref{Fig5}. Three simulations with different values of $R_c/\lambda_L$ are performed for each value of $\Gamma$. In all simulations, the target is assumed to be irradiated by a planar circularly polarized laser wave, and the following target parameters are used: radius $R_0 = 1.14 R_c$, maximum thickness $d_0=1.2 d_{\rm opt}$, and plasma density $n_0 = 100n_{\rm c}$ (see supplemental material for the reasons of choosing these parameters).

\begin{table}[ht]
\caption{\label{table1}Simulation parameters}
\begin{ruledtabular}
\begin{tabular}{clcccc}
\ &$\Gamma^{1/2}({\rm PW\footnote{laser power in petawatt}})$&0.67(2.8)&0.87(8)&1.07(18)&1.27(35)\\
\hline
$\bf{x}$        &$R_c/\lambda_L$&1.9    &2.6     &3.4    &4.1\\
$\bullet$       &$R_c/\lambda_L$&3.2    &4.5     &5.8    &7.0\\
$\blacktriangle$&$R_c/\lambda_L$&6.2    &7.5     &8.8    &10\\
\end{tabular}
\end{ruledtabular}
\end{table}

The identification of the stable LILA regimes was done by analyzing the transverse size of the hot spot at the focal point, as well as the particle/energy densities within it. For example, the simulation results shown in Fig.~\ref{Fig3} corresponding to $\Gamma=1.6$ and $R_c/\lambda_L = 7$ exemplify a stable focusing case. Strong convergence of the target appears to suppress the instability. In fact, one of the characteristic signatures of the RT instability is the breakup of the target into multiple density clumps. Such instability onset is indeed observed at $t\approx 26.6$fs. However, the clumps converge towards the axis and merge at the later times, thereby effectively suppressing the instability. 

On the contrary, Fig.~\ref{Fig5}~(d) shows a typical example of {\it unstable} target focusing corresponding to $\Gamma=1.61$ and $R_c/\lambda_L = 10$. The RT instability breaks the target into large density clumps, the RPA fails to focus the target, and the acceleration eventually terminates because the entire target becomes transparent to laser light after $t=66$fs. As a result, the target is dispersed at $t=133$fs. When compared with the focused target in Fig.~\ref{Fig3}, the peak ion density of the RT-unstable target is reduced by one order of magnitude, and the ion energy spectrum shown in Fig.~\ref{Fig5}~(e) is no longer mono-energetic.

One immediate observation from Figure~\ref{Fig5}~(a) is that, for given laser power, the target focusing is stabilized at small values of $R_c$, but is disrupted for larger targets. Qualitatively, this can be understood by observing that larger $R_c$ corresponds to longer focusing time, thus supporting more e-foldings for the developing RT-like instability.  Figure~\ref{Fig5}(a) further implies that, for a given target size, its focusing/acceleration is stabilized for large values of $\Gamma$. This result is consistent with earlier calculations~\cite{pegoraro2007RPA_2,qiao2009RPA_4}: higher laser power accelerates ions to higher velocities and, therefore, provides less time for the instability to grow. Not surprisingly, whenever the conditions for stable acceleration/focusing are met, the predictions of the DM model for the focal length $f$ and the ion momentum $p_x$ are very accurate. Indeed, the results obtained with VLPL simulations are in agreement with Eq.(\ref{eq_f}) as shown in Figures~\ref{Fig5}(b,c). The ion momentum $p_x$ obtained at the maximum of the distribution function is found to be close to the momentum average over all ions inside the hot spot.

{\it Discussion and outlook.} Simultaneous acceleration and focusing of the variable-thickness LILA targets have been shown to be stable under various laser powers, ranging from $2.8PW$ to $35PW$. Near the lower end of this power range, ultrashort circularly polarized lasers are already available~\cite{Korea_4PW,CP_Polymer_PW}. The $10$s-PW laser pulses will soon become accessible at several user facilities worldwide~\cite{mourou2011eli_whitebook,Apollon_10PW,Shanghai_10PW}. Not only are the LILA targets stable in different power regimes, but they are also robust under various laser-target configurations. According to the DM model, realistic LILA targets composed of proton-rich materials (e.g., F8BT polymers~\cite{CP_Polymer_PW}) should behave similarly to pure hydrogen targets as long as the dimensionless $\Gamma$ parameter is rescaled to account for the changed effective $Z/M$ ratio of a multi-species target. This conclusion is also confirmed by a 3D-PIC simulation (see the supplemental material) of a C-H LILA target. 

Furthermore, LILA targets can be successfully focused by realistic laser pulses with non-planar transverse profiles. This is accomplished by correcting the target thickness profile $d(r_0)$ according to the transverse profile of the pulse. For a Gaussian laser pulse ($I=I_0 \exp(-r^2/\sigma_L^2)$), the DM model predicts that the optimal target thickness profile must be corrected according to $d(r_0)\rightarrow d(r_0)\exp(-r^2/\sigma_L^2)$. Our 3D-PIC simulations (see supplemental material) support this conclusion. All simulations presented in this Letter assumed a flat-top longitudinal laser profile with a $\tau_{\rm rise} = 3.3$fs rising edge. Current laser technology is steadily progressing towards high-contrast PW laser pulse with contrasts well above $10^{12}$ \cite{Contrast_1,Contrast_2}, thereby avoiding deleterious pre-plasma generation. Advances in laser and nanofabrication technologies will enable the experimental realization of our theoretical concept of simultaneous acceleration and focusing ion beams by ultra-intense laser pulses. The scaling laws presented here enable designing the target geometry and selecting the appropriate laser power and duration. Depending on those parameters, a wide range of ion kinetic energies -- from $200$MeV to $750$MeV -- can be obtained, with future applications in sight: ion accelerators for cancer treatment \cite{schardt2010heavy_proton_therapy}, novel spallation sources~\cite{mckenna2005spallation_1}, and many others.


\section{ACKNOWLEDGMENT}
This work is supported by the award from the HEDLP NNSA Grant NA0003879. The authors thank the Texas Advanced Computing Center (TACC) at The University of Texas at Austin for providing HPC resources.


\begin{thebibliography}{10}



\bibitem{GRIN_Book}
    R. G. Griggers,
    "Encyclopedia of Optical Engineering,"
    {\em CRC Press}, vol~1, 675 (2003).


\bibitem{macchi2005RPA_1}
    A.~Macchi, F.~Cattani, T.~V.~Liseykina, and F.~Cornolti,
    "Laser Acceleration of Ion Bunches at the Front Surface of Overdense Plasmas,"
    {\em Phys. Rev. Lett.}, vol~94, 165003 (2005).

\bibitem{pegoraro2007RPA_2}
    F.~Pegoraro and S.~V.~Bulanov,
    "Photon Bubbles and Ion Acceleration in a Plasma Dominated by the Radiation Pressure of an Electromagnetic Pulse,"
    {\em Phys. Rev. Lett.}, vol~99, 065002 (2007).

\bibitem{robinson2008RPA_3}
    A.~Robinson, M.~Zepf, S.~Kar, R.~Evans, and C.~Bellei,
    "Radiation Pressure Acceleration of Thin Foils with Circularly Polarized Laser Pulses,"
    {\em New J. Phys.}, vol~10, 013021 (2008).


\bibitem{qiao2009RPA_4}
    B. Qiao, M. Zepf, M. Borghesi, and M. Geissler,
    "Stable GeV Ion-Beam Acceleration from Thin Foils by Circularly Polarized Laser Pulses,"
    {\em Phys. Rev. Lett.}, vol~102, 145002 (2009).

\bibitem{fourkal2005coulomb}
    E. Fourkal, I. Velchev, and C.-M. Ma,
    "Coulomb Explosion Effect and the Maximum Energy of Protons Accelerated by High-power Lasers,"
    {\em Phys. Rev. E}, vol~71, 036412 (2005).


\bibitem{pukhov_prl10}
    Tong-Pu Yu, Alexander Pukhov, Gennady Shvets, and Min Chen,
    "Stable Laser-Driven Proton Beam Acceleration from a Two-Ion-Species Ultrathin Foil,"
    {\em Phys. Rev. Lett.} vol~105, 065002 (2010).

\bibitem{khudik2014analytic_1D}
    V. Khudik, S. Yi, C. Siemon, and G. Shvets,
    "The Analytic Model of a Laser-accelerated Plasma Target and Its Stability,"
    {\em Phys. Plasmas}, vol~21, 013110 (2014).

\bibitem{tabak1994ignition}
    M. Tabak, J. Hammer, M. E. Glinsky, W. L. Kruer, S. C. Wilks, J. Woodworth, E. M. Campbell, M. D. Perry, and R. J. Mason,
    "Ignition and High Gain with Ultrapowerful Lasers,"
    {\em Phys. Plasmas}, vol~1, 1626 (1994).

\bibitem{atzeni2002ignition_2}
    S. Atzeni, M. Temporal, and J. Honrubia,
    "A first analysis of fast ignition of precompressed ICF fuel by laser-accelerated protons,"
    {\em Nucl. Fusion}, vol~42, L1 (2002).

\bibitem{patel2003WDM_1}
    P. Patel, A. Mackinnon, M. Key, T. Cowan, M. Foord, M. Allen, D. Price, H. Ruhl, P. Springer, and R. Stephens,
    "Isochoric Heating of Solid-Density Matter with an Ultrafast Proton Beam,"
    {\em Phys. Rev. Lett.} vol~91, 125004 (2003).
\bibitem{dyer2008WDM_2}
    G.~.M.~Dyer, A. C. Bernstein, B. I. Cho, J. Osterholz, W. Grigsby, A. Dalton, R. Shepherd, Y. Ping, H. Chen, K. Widmann and T. Ditmire,
    "Equation-of-State Measurement of Dense Plasmas Heated With Fast Protons,"
    {\em Phys. Rev. Lett.} vol~101, 015002 (2008).

\bibitem{bulanov2002therapy_1}
    S. Bulanov, T. Z. Esirkepov, V. Khoroshkov, A. Kuznetsov, and F. Pegoraro,
    "Oncological Hadrontherapy with Laser Ion Accelerators,"
    {\em Phys. Lett. A} vol~299, 240 (2002).

\bibitem{yogo2009therapy_2}
    A. Yogo et al.,
    "Application of Laser-accelerated Protons to the Demonstration of DNA Double-strand Breaks in Human Cancer Cells,"
    {\em Appl. Phys. Lett.} vol~94, 181502 (2009).

\bibitem{bolton2010therapy_3}
    P. Bolton, T. Hori, H. Kiriyama, M. Mori, H. Sakaki,K. Sutherland, M. Suzuki, J. Wu, and A. Yogo,
    "Toward Integrated Laser-driven Ion Accelerator Systems at the Photo-medical Research Center in Japan,"
    {\em Nucl. Instrum. Meth. Phys. Res. A} vol~620, 71 (2010).

\bibitem{mckenna2005spallation_1}
    P. McKenna, K. W. D. Ledingham, S. Shimizu, J. M. Yang, L. Robson, T. McCanny, J. Galy, J. Magill, R. J. Clarke, D. Neely, P. A. Norreys, R. P. Singhal, K. Krushelnick, and M. S. Wei,
    "Broad Energy Spectrum of Laser-Accelerated Protons for Spallation-Related Physics,"
    {\em Phys. Rev. Lett.} vol~94, 084801 (2005).

\bibitem{hannachi2007prospects_nuclear}
    F. Hannachi, M. Aléonard, M. Gerbaux, F. Gobet, G. Malka, C. Plaisir, J. Scheurer, M. Tarisien, P. Audebert, E. Brambrink, V. Méot, P. Morel, Ph. Nicolaï and V. Tikhonchuk,
    "Prospects for Nuclear Physics with Lasers,"
    {\em Plasma Phys. Control. Fusion} vol~49, B79 (2007).

\bibitem{perry1999petawatt}
    M. Perry, D. Pennington, B. Stuart, G. Tietbohl, J. Britten, C. Brown, S. Herman, B. Golick, M. Kartz, J. Miller,  H. T. Powell, M. Vergino, and V. Yanovsky,
    "Petawatt Laser Pulses,"
    {\em Opt. Lett.} vol~24, 160 (1999).

\bibitem{esirkepov2004Model_1D_efficiency}
    T. Esirkepov, M. Borghesi, S. V. Bulanov, G. Mourou, and T. Tajima,
    "Highly Efficient Relativistic-Ion Generation in the Laser-Piston Regime,"
    {\em Phys. Rev. Lett.} vol~92, 175003 (2004).


\bibitem{bulanov2010unlimitedModel_1D}
    S. V. Bulanov, E. Y. Echkina, T. Z. Esirkepov, I. N. Inovenkov, M. Kando, F. Pegoraro, and G. Korn,
    "Unlimited Ion Acceleration by Radiation Pressure,"
    {\em Phys. Rev. Lett.} vol~104, 135003 (2010).

\bibitem{yan2008spectrum_1}
    X. Q. Yan, C. Lin, Z.-M. Sheng, Z. Y. Guo, B. C. Liu, Y. R. Lu, J. X. Fang, J. E. Chen,
    "Generating High-Current Monoenergetic Proton Beams by a CircularlyPolarized Laser Pulse in the Phase-StableAcceleration Regime,"
    {\em Phys. Rev. Lett.} vol~100, 135003 (2008).

\bibitem{snavely2000TNSA_1}
    R. A. Snavely, M. H. Key, S. P. Hatchett, T. E. Cowan, M. Roth, T. W. Phillips, M. A. Stoyer, E. A. Henry, T. C. Sangster, M. S. Singh, S. C. Wilks, A. MacKinnon, A. Offenberger, D. M. Pennington, K. Yasuike, A. B. Langdon, B. F. Lasinski, J. Johnson, M. D. Perry, and E. M. Campbell,
    "Intense High-Energy Proton Beams from Petawatt-Laser Irradiation of Solids,"
    {\em Phys. Rev. Lett.} vol~85, 2945 (2000).

\bibitem{hatchett2000TNSA_2}
    S. P. Hatchett et al.,
    "Electron, Photon, and Ion Beams from the Relativistic Interaction of Petawatt Laser Pulses with Solid Targets,"
    {\em Phys. Plasmas} vol~7, 2076 (2000).

\bibitem{silva2004shocks_1}
    L. O. Silva, M. Marti, J. R. Davies, R. A. Fonseca, C. Ren, F. S. Tsung, and W. B. Mori,
    "Proton Shock Acceleration in Laser-Plasma Interactions,"
    {\em Phys. Rev. Lett.} vol~92, 015002 (2004).

\bibitem{PhysRevLett.101.164802shocks_2}
    L. Ji, B. Shen, X. Zhang, F. Wang, Z. Jin, X. Li, M. Wen, and J. R. Cary,
    "Generating Monoenergetic Heavy-Ion Bunches with Laser-Induced Electrostatic Shocks,"
    {\em Phys. Rev. Lett.} vol~101, 164802 (2008).

\bibitem{yin2006BOA_1}
    L. Yin, B. Albright, B. Hegelich, and J. Fernández,
    "GeV Laser Ion Acceleration from Ultrathin Targets: The Laser Break-out Afterburner,"
    {\em Laser Part. Beams} vol~24, 291 (2006).

\bibitem{BrunleHeating}
    F. Brunel,
    "Not-so-resonant, Resonant Absorption,"
    {\em Phys. Rev. Lett.} vol~59, 52 (1987).

\bibitem{BrunleHeating2}
    F. Brunel,
    "Anomalous Absorption of High Intensity Subpicosecond Laser Pulses,"
    {\em Phys. Fluids} vol~31, 2714 (1988).


\bibitem{munson2014fundamentals}
    B. R. Munson, T. H. Okiishi, A. P. Rothmayer, and W. W. Huebsch,
    "Fundamentals of Fluid Mechanics,"
    {\em John Wiley \& Sons} (2014).

\bibitem{gjurchinovski2004reflection}
    A. Gjurchinovski,
    "Reflection of Light from a Uniformly Moving Mirror,"
    {\em Am. J. Phys.} vol~72, 1316 (2004).

\bibitem{galli2012reflection}
    J. R. Galli and F. Amiri,
    "A General Principle for Light Reflecting from a Uniformly Moving Mirror: A Relativistic Treatment,"
    {\em Am. J. Phys.} vol~80, 680 (2012).

\bibitem{hecht2016optics}
    E.~Hecht,
    "Optics,"
    {\em Pearson Education}, (2016).


\bibitem{pukhov1999VLPL3D}
    A. Pukhov,
    "Three-dimensional Electromagnetic Relativistic Particle-in-cell Code VLPL (Virtual Laser Plasma Lab),"
    {\em J. Plasma Phys.} vol~61, 425 (1999).


\bibitem{tripathi2009_optimal_thick}
    V. Tripathi, C.-S. Liu, X. Shao, B. Eliasson, and R. Z. Sagdeev,
    "Laser Acceleration of Monoenergetic Protons in a Self-organized Double Layer from Thin Foil,"
    {\em Plasma Phys. Control. Fusion} vol~51, 024014 (2009).



\bibitem{Korea_4PW}
    J.~H.~Sung, H.~W.~Lee, J.~Y.~Yoo, J.~W.~Yoon, C.~W.~Lee, J.~M.~Yang, Y.~J.~Son, Y.~H.~Jang, S.~K.~Lee, and C.~H.~Nam,
    "4.2 PW, 20 fs Ti: sapphire laser at 0.1 Hz,"
    {\em Opt. Lett.} vol~42, 11 (2017).
\bibitem{CP_Polymer_PW}
    I.~J.~Kim, K.~H.~Pae, I.~W.~Choi, CL.~Lee, H.T.~Kim, H.~Singhal, J.~H.~Sung , S.~K.~Lee, H.~W.~Lee, P.~V.~Nickles, T.~M.~Jeong, C.~M.~Kim, and C.~H.~Nam,
    "Radiation pressure acceleration of protons to 93 MeV with circularly polarized petawatt laser pulses,"
    {\em Phys. Plasmas} vol~23, 7 (2016).
\bibitem{mourou2011eli_whitebook}
    G. A. Mourou, G. Korn, W. Sandner, and J. L. Collier,
    "ELI WHITEBOOK,"
    {\em THOSS Media GmbH} (2011).
\bibitem{Apollon_10PW}

    B.~Le Garrec, D.~N.~Papadopoulos, C.~Le~Blanc, J.~P.~Zou, G.~Chériaux, P.~Georges, F.~Druon, L.~Martin, L.~Fréneaux, A.~Beluze, N.~Lebas; F.~Mathieu, P.~Audebert,
    "Design update and recent results of the Apollon 10 PW facility,"
    {\em Proc. SPIE} vol~10238, 80 (2017).
\bibitem{Shanghai_10PW}
    W.~Li, Z.~Gan, L.~Yu, C.~Wang, Y.~Liu, Z.~Guo, L.~Xu, M.~Xu, Y.~Hang, Y.~Xu, J.~Wang, P.~Huang, H.~Cao, B.~Yao, X.~Zhang, L.~Chen, Y.~Tang, S.~Li, X.~Liu, S.~Li, M.~He, D.~Yin, X.~Liang, Y.~Leng, R.~Li, and Z.~Xu,
    "339 J high-energy Ti: sapphire chirped-pulse amplifier for 10 PW laser facility,"
    {\em Opt, Lett.} vol~43, 22 (2018).
\bibitem{Contrast_1}
    J.~M.~Mikhailova, A.~Buck, A.~Borot, K.~Schmid, C.~Sears, G.~D.~Tsakiris, F.~Krausz, and L.~Veisz,
    "Ultra-high-contrast few-cycle pulses for multipetawatt-class laser technology,"
    {\em Opt. Lett.} vol~36, 16 (2011).
\bibitem{Contrast_2}
    Y.~Wang, S.~Wang, A.~Rockwood, B.~M.~Luther, R.~Hollinger, A.~Curtis, C.~Calvi, C.~S.~Menoni, and J.~J.~Rocca,
    "0.85 PW laser operation at 3.3 Hz and high-contrast ultrahigh-intensity $\lambda$ = 400 nm second-harmonic beamline,"
    {\em Opt. Lett.} vol~42, 19 (2017).


\bibitem{schardt2010heavy_proton_therapy}
    D. Schardt, T. Elsasser, and D. Schulz-Ertner,
    "Heavy-ion Tumor Therapy: Physical and Radiobiological Benefits,"
    {\em Rev. Mod. Phys} vol~82, 383 (2010).


\end{thebibliography}
\end{document}